\documentstyle [prl,aps]{revtex}
\input epsf

\begin{document}

\title {A New Solar System Population of WIMP Dark Matter}

\author{Thibault Damour$^{1}$ and Lawrence M. Krauss$^2$}
\address{
 $^1$Institut des Hautes Etudes Scientifiques, 91440
Bures-sur-Yvette,  France\\
and DARC, Observatoire de Paris-CNRS, F-92195 Meudon, France\\
$^2$Departments of Physics and Astronomy
Case Western Reserve University
Cleveland, OH~~44106-7079
}

\twocolumn[
\maketitle
\begin{abstract}
\widetext
\noindent Perturbations due to the planets combined with the
non-Coulomb nature of the gravitational potential in the Sun imply
that WIMPs that are gravitationally captured by scattering in surface layers of
the Sun can evolve into orbits that no longer intersect the Sun.
For orbits with a semi-major axis $ < 1/2$ of Jupiter's orbit, such WIMPs
can persist in the solar system for $ > 10^9$ years, leading to a previously
unanticipated population intersecting Earth's orbit.   For
WIMPs detectable in the next generation of detectors, this population can
provide a complementary signal, in the keV range, to that of galactic halo dark
matter.
\end{abstract}


]

\narrowtext

The effort to uncover the nature of the dark matter that dominates the
gravitational potential well in our galaxy, and almost all known galaxies, is
perhaps one of the most important ongoing experimental programs in cosmology.
Among the favored candidates for non-baryonic dark matter are so called 
Weakly Interacting Massive Particles (WIMPs).  Such WIMPs naturally arise in
most supersymmetric extensions of the Standard Model 
in the form of  the lightest
supersymmetric partner of normal matter, the 
neutralino  (e.g. \cite{jungetal}).
A number of ongoing
experiments, involving underground low background detectors, capable of
detecting the energy deposited due to the elastic scattering of neutralinos 
off atoms, are underway. The tip
of the allowed parameter space of neutralinos
 is just beginning to be probed.

If a positive detection is made in the next generation of underground
detectors, it will be important to search for signatures that
definitively distinguish such a signal from the possible background noise due to
radioactive contaminants in and around the detectors.   Possibilities include
searches for a 
potential annual modulation and/or angular anisotropy of the
signal\cite{sperg,krausscop,jungetal}, or indirect searches involving the
annihilation products of the neutralino populations captured by the Sun
\cite{kraussetal,Srednicki,Gould,jungetal} 
or the Earth \cite{krauswilcsred,gould}.

The fact that WIMPs can be captured by the Sun and planets motivates
examining whether various complicated dynamical histories in the solar system
might affect the local WIMP density near the Earth (e.g.
\cite{gouldind,goulddif}) As we describe here, a careful consideration of the
WIMPs that are scattered by a nucleus in the Sun into gravitationally bound
orbits indicates that a small population can elude subsequent scattering in the
solar interior (which otherwise leads to concentration in the Solar core,
followed by annihilation).  Because of the non-Coulomb nature of the
gravitational potential in the Sun, this population involves only WIMPs that
scatter near the surface of the Sun.  These WIMPs can, due to the perturbing
influence of the planets, diffuse out of the Sun and build up over
time to produce a density in the region of the Earth that may be comparable to
the background WIMP halo density. The range of scattering cross sections and
masses for such WIMPs is precisely that which is associated with the range of
parameter space just below current experimental limits.  Thus, if WIMPs are to 
be
discovered in the next  generation of experiments, then the population we
describe here should be significant.

We shall focus on the sub-population of WIMPs that scatter on a nucleus
located near the surface of 
the Sun, and  thereby lose just enough energy to stay in Earth-crossing
orbits.  These will be susceptible to small
gravitational perturbations by the planets.  We are interested then in
the differential capture rate, per energy, and per angular momentum, of
WIMPs in the Sun, and in particular only in the fraction of WIMPs that
have angular momenta in a small range 
$[J_{\rm min}, J_S]$ 
where $J_S$ is the angular momentum for a WIMP exactly grazing the Sun.
The standard low-energy
differential cross section of WIMPs on nuclei of atomic number $A$ is given
by $
d \, \sigma_A = \sigma_A^0 \, F_A^2 (Q) \, \frac{d \, \Omega_{\rm 
cm}}{4\pi} $
where $Q = E_{\rm before} - E_{\rm after}$ is the energy transferred 
during the scattering, $F_A (Q)$ is a form factor, 
and $d
\Omega_{\rm cm} $ is the  scattering solid angle element {\it in the center of
mass} frame.  Note 
that $\sigma_A^0$  is by 
definition independent of the scattering angle, and takes a value that 
depends on the details of the WIMP particle physics parameters.  
Generalizing the standard \cite{gould} capture calculation 
(see \cite{damkrauss}), one can derive 
the rate with 
which WIMPs scatter on nuclei with atomic number $A$ to end up into bound 
solar orbits with semi-major axis between 
$[a , a + da]$ (corresponding to 
$[\alpha , \alpha + d \alpha]$ with $\alpha \equiv G_N \, m_{\rm Sun} / a$), and
with  specific angular momentum $J_{\rm min} \leq J \leq J_{S}$:

\begin{eqnarray}
&& \left.\frac{d \dot{N}_A}{d \, \alpha} \right\vert_{J \geq J_{\min}} 
\simeq \frac{n_X}{v_o}  \nonumber \int_{r \geq r_{\min}} d^3 r  n_A (r) 
\sigma_A^0 \\ 
&&  \times  \left( 1 - \frac{J_{\min}^2}{r^2 \, [v_{\rm esc}^2 (r)-\alpha]} 
\right)^{1/2} K_A (r,\alpha) . \label{eq2.20}
\end{eqnarray}

 Here, the minimum radius
$r_{\min}$  (perihelion) is defined in terms of the minimum angular
momentum 
$J_{\min}$ by $r_{\min} \, \sqrt{v_{\rm esc} (r_{\min})^2 - \alpha} 
\equiv J_{\min}$,
where $v_{\rm esc} (r_{\min})$ is the escape velocity at $r_{\min}$,  
$n_{A,X}$ is the density of atomic targets $A$, and WIMPs $X$,
respectively, $v_o$ is the rms circular velocity of WIMPs in
the Galaxy, and the ``capture function"
$K_A (r,\alpha)$ involves an integral over the WIMP local
phase space distribution at the Sun, weighted over the
scattering form factor \cite{damkrauss}.  While the general form (used in
deriving our results) is not particularly illuminating, in the limit in which 
the
nuclear form factor is neglected in the scattering cross section, and in which 
we
also neglect the relative motion of the Sun with respect to the galactic
halo, $K_A (r,\alpha)$ takes the following simple form: 
\begin{eqnarray}
&& K_A (r,\alpha) = \frac{2}{\pi^{1/2}} \, 
\frac{1}{\beta_+^A}  \nonumber \\
&& \times \left[ 1 - \exp \left[ - \frac{\beta_-^A}{v_o^2} 
\, \left( v_{\rm esc}^2 (r) - \frac{\alpha}{\beta_+^A} \right) \right] 
\right] \, . \label{eq2.26}
\end{eqnarray}
where 
$\beta_{\pm}^A \equiv \frac{4 \, m_X \, m_A}{(m_X \pm m_A)^2} \,$.

To obtain the total rate, a sum over all the (significant) values of $A$
present in the Sun must be ultimately performed. 
We shall be 
interested in values $J_{\min} \simeq J_S =R_S v_{\rm esc} (R_S)$,
where $R_S$ is the Sun's
radius, and
 $a \sim 1 \, {\rm AU}$, i.e. $\alpha \sim Gm_{\rm Sun} / (1 \,
{\rm AU}) \sim v_E^2$  where $v_E = 29.8 
\, {\rm km/s}$ is the Earth orbital velocity. Note that for such values 
of $a$, typically $\alpha \sim v_E^2 \ll v_{\rm esc}^2$, and from the above
formula it is clear that 
the  function $K_A (r,\alpha)$ is nearly independent of $\alpha$. 

The Sun scattering events create a 
population of solar-system bound WIMPs, moving (for $ a \sim 1 \, {\rm 
AU}$) on 
very elliptic orbits that traverse the Sun over and over again. For the 
values of WIMP-nuclei cross sections we shall be mostly interested in 
here (corresponding to effective WIMP-proton cross sections in the range 
$4 \times 10^{-42} - 4 \times 10^{-41} \, {\rm cm}^2$), the mean opacity 
of the Sun for orbits with small perihelions is in the range 
$10^{-4} - 10^{-3}$. This means that after only $10^3 - 10^4$ orbits 
these WIMPs will undergo a second 
scattering event in the Sun.  It is straightforward to show that
this second scattering event significantly reduces the semi-major axis of
the WIMP, removing them from the population of interest here (They will end up
in the core of the Sun where they  will ultimately annihilate with
each other). 

The only  way to save some of these WIMPs from this 
demise is to consider a fraction of WIMPs that have perihelions
$r_{\rm min}$ in a small  range near the radius of the Sun, say $R_S
(1-\epsilon) \leq r_{\rm  min} \leq R_S$. Focussing on such a subpopulation of
WIMPs has two  advantages: (i) they traverse a small fraction of the mass
of the Sun and therefore their lifetime on such grazing orbits is greatly
increased, and, {\it more importantly} (ii)  during this time, gravitational
perturbations due to the planets can push them into orbits that no
longer cross the Sun. 

We first recall some 
concepts and notation of Hamiltonian dynamics. In standard 
position-momenta variables the Hamiltonian describing the basic 
interaction between a WIMP and the Sun reads (for $r \equiv \vert {\bf x} 
\vert)$,
 ${\cal H}_S ({\bf x} , {\bf p}) = \frac{1}{2} \, {\bf p}^2 - U(r) \, , 
$
where $U(r) = + \, G_N \int \rho ({\bf x}') \, d^3 {\bf x}' / \vert {\bf x} - 
{\bf x}' \vert$ is the (spherically symmetric) Newtonian potential 
generated by the mass distribution of the Sun. Note that the mass $m_X$ of the
WIMP has been factored out of
all quantities. We work with action-angle variables (``Delaunay variables''), 
traditionally denoted $(L,G,H ; \ell ,g,h)$ \cite{brouwer}. The action 
variables $L$, $G$, $H$ are related to $E$, $J$ and $J_z$, respectively.
 The angle variables (with period $2\pi$) corresponding to $L$, $G$, $H$ are 
respectively denoted $\ell$, $g$, $h$. In these variables the Hamiltonian
depends only on $L$  and $G$ and the general evolution equations,

\label{eq2}
\begin{eqnarray}
\frac{d\ell}{dt} & = & + \frac{\partial {\cal H}}{\partial L} \ , \ 
\frac{dg}{dt} = + \frac{\partial {\cal H}}{\partial G} \ , \ 
\frac{dh}{dt} = + \frac{\partial {\cal H}}{\partial H} \, , 
\label{eq4} \\
&& \nonumber \\
\frac{dL}{dt} & = & - \frac{\partial {\cal H}}{\partial \ell} \ , \ 
\frac{dG}{dt} = - \frac{\partial {\cal H}}{\partial g} \ , \ 
\frac{dH}{dt} = - \frac{\partial {\cal H}}{\partial h} \, , 
\label{eq5}
\end{eqnarray}
tell us, in the case of the Hamiltonian above, that the action 
variables $L$, $G$ and $H$ are constant, while, among the angle 
variables, $h$ is constant, but $\ell$ and $g$ evolve linearly in time: 
$\ell = n t + \ell_o$, $g = \dot{\omega} \, t + g_o$. Here, $n \equiv 
2\pi / P$  is the mean angular frequency of the radial motion $(P =$ 
perihelion to perihelion period), and $\dot \omega$ is the mean rate of 
advance of the perihelion.

The crucial point to realize is the following. If we consider a WIMP orbit 
with a generic perihelion $r_{\rm min} < R_S$, it will 
undergo a large perihelion precession $\Delta \, \omega \sim 2\pi$ per orbit, 
i.e. $\dot \omega \sim n$, because the potential $U(r)$ {\it within the 
Sun} is modified compared to the exterior $1/r$ potential leading  to the
absence of perihelion motion. In other words, the  trajectory of the WIMP will
generically be a fast advancing {\it rosette}.  This means that {\it both}
angles $\ell$ and $g$ are {\it fast variables}.  When adding in the small
perturbing effect due to the planets, i.e. when  considering the total
Hamiltonian,
$ {\cal H}_{\rm tot} = {\cal H}_S (L,G) + {\cal H}_p (L,G,H ; \ell ,g,h ; 
L_p , \ldots , \ell_p , \ldots )$
where ${\cal H}_p$ (which contains a small factor $\mu_p = m_{\rm planet} 
/ m_S$) is the planetary perturbation, we can work out the (first-order) 
{\it secular} effects due to the planets by averaging over the fast 
variables $\ell$ and $g$ (as well as the mean anomalies $\ell_p$ of the 
planets). Then the evolution equations tell us immediately 
that the corresponding action variables $L$ and $G$ are secularly constant 
because planetary perturbations average out to zero (e.g. $\langle dG / dt 
\rangle = - \langle \partial {\cal H} / \partial g \rangle_{\ell , g} 
\equiv 0$).  $L$ is
essentially related to the semi-major axis $a$ of the WIMP  orbit, while $G/L$
is related to the eccentricity $e$. As a result, when the rosette
motion is fast, planetary perturbations do not induce any secular evolution in
the semi-major axis and in the  eccentricity of the WIMP orbit. Such
WIMPs will end up
in the  core of the Sun.

A new situation arises for WIMP orbits that graze the Sun, because these 
orbits feel essentially a $1/r$ potential due to the Sun, so that their 
rosette motion will be very slow. Consequently, the variable $g$ will be 
slow (compared to $\ell$), and we cannot average over $g$.
We can split the total Hamiltonian in 
three parts ${\cal H}_{\rm tot} = {\cal H}_o + {\cal H}_1 + {\cal H}_p $
where we take as unperturbed Hamiltonian the one corresponding to a 
point-like Sun
while the perturbations are
${\cal H}_1 = -\delta \, U(r) $ and $ {\cal H}_p = \sum_p -G_N  m_p \left( 
\frac{1}{\vert {\bf x}_X - {\bf x}_p \vert} - \frac{{\bf x}_X \cdot {\bf 
x}_p}{\vert {\bf x}_p \vert^3} \right) $.
Here, $\delta \, U(r) \equiv U(r) - \ G_N \, m_S / r$ is the non-$1/r$ 
part of the potential generated by the Sun, and ${\cal H}_p$ denotes the
planetary perturbations involving a sum over the planets with
masses
$m_p$ and heliocentric positions 
${\bf x}_p$. [The last term comes from the transformation between inertial 
(barycentric) coordinates and heliocentric ones.]

Using Delaunay variables defined by the
Hamiltonian ${\cal H}_o$, we can derive the {\it secular} evolution of 
 $L$, $G$, $H$, under the combined influence of the 
perturbations ${\cal H}_1$ and ${\cal H}_p$. This is simply obtained by 
averaging the canonical equations over the fast angles, i.e. all the mean 
anomalies of the problem: $\ell$, $\ell_p$. [We denote this average by an 
overbar.] By averaging the evolution equations  one finds that (in first 
order) $L$ will be secularly constant (i.e. $a =$ const), while $G$, $H$, 
$g$, $h$ slowly evolve under the averaged perturbed Hamiltonian $
{\cal H}_{\rm pert} (L,G,H;g) = \overline{\cal H}_1 (L,G) + \overline{\cal 
H}_p (G,H;g;L_p) $.

Because ${\cal H}_{\rm pert}$ does not depend on the angle $h$, its 
conjugate momentum $H =J_z$ will be secularly constant. Determining the secular
evolution of the remaining canonical pair $(G,g)$, and therefore estimating
the minimum values of the WIMP angular momenta 
$J_{\min} \equiv G_{\min}$ for which planetary perturbations are strong 
enough to kick them out of the Sun is then reduced to studying the level 
curves of ${\cal H}_{\rm pert} (G;g)$. 
Among these level curves, there exists a {\it separatrix} $\cal S$
 such that initial orbits ``above'' $\cal S$ secularly evolve into orbits
with angular momenta $G > G_S$, i.e. large enough to no longer intersect the
Sun.
 We find
that it 
generically takes (for $a \sim 1 {\rm AU}$) less than $10^3$ WIMP radial 
periods (i.e. less than $10^3 \, {\rm yr}$) for the eccentricity of a 
WIMP above $\cal S$
 to increase sufficiently to exit the Sun. Then, once $G > G_S$ 
the time scale for the subsequent evolution of $G$ is
given by the planetary  perturbations alone and is roughly $1/[
{\sum_{p}}\mu_p (a/a_p)^3 ]$ longer than one orbital period, i.e. roughly 
$10^5 \, {\rm yr}$ for $a \sim a_1 \equiv 1 \, {\rm AU}$. After this time, 
the WIMP would, if it evolved only under the simplified planetary 
Hamiltonian $\overline{\cal H}_p$, come back again to low values of the 
eccentricity, corresponding to Sun-penetrating orbits. Under the influence 
of $\overline{\cal H}_1$, it would then again bounce back away from the 
Sun in $\sim 10^3$ orbits. For the scattering cross sections we shall be 
discussing below, the opacity of the small outer skin of the Sun is
typically smaller than $\sim 10^{-5}$. Therefore the  above process could
persist for hundreds of cycles before the WIMP gets scattered again in
the Sun. However, it is clear  that the real
gravitational interaction of the WIMP with planets is much  more
complicated than what is described by $\overline{\cal H}_p$. In 
particular, the non zero eccentricities of the other planets, and the 
occurrence, once in a while, of a near collision with an inner planet
will  cause the elliptic elements of the WIMP to diffuse chaotically away
from  the simplified periodic history described above. As the very high 
eccentricities (for AU-size orbits) needed to traverse the Sun  represent
only a very small fraction of the phase space into which the  WIMP can
diffuse, on time scales of a few million  years most of the population of
WIMPs we are talking about will have  irreversibly evolved onto
trajectories that do not intersect the Sun for the entire age of the
solar  system. We further estimate that the long-term survival  of such
WIMPs on orbits that stay within the inner solar system is  greater than
4.5 Gyr if $ a < a_{\rm Jup}/2 =2.6 a_E$. [This crucial assertion
is based on an approximate analytical estimate
of the lifetime of WIMPs \cite{damkrauss} which
should be checked by dedicated numerical 
simulations.]

We can finally estimate the density of WIMPs that diffuse out to solar-bound
orbits and which can survive to the present time by simply integrating our
differential capture rate over all trajectories $J > J_{\min} \equiv G_{\min}$
that end up out of the Sun, suitably averaged over the initial WIMP
distribution.  The rate (per $\alpha = G_N \, m_S / a$) of solar capture of
WIMPs that  subsequently survive out of the Sun to stay within the inner solar
system then depends on the $A$-dependent combination:
$ g_A \equiv \frac{f_A}{m_A} \, \sigma_A^0 \, K_A^s $, where $f_A$ is the
fraction (by mass) of element $A$ in the Sun, and $ K_A^s $ is the
Sun-surface value of the capture function mentioned above.
Note that the $A$-dependence is entirely contained in  $g_A$ 
with dimensions $[\hbox{cross section}] / [\hbox{mass}]$.  The total capture
rate is then dependent on ${\displaystyle \sum_A} 
\, g_A  =g_{\rm tot}$.

By integrating this capture rate, and making
simplifying assumptions about the orbital evolution
of the considered WIMPs, we can estimate both 
the present space and velocity distributions of this new WIMP population. 
 We find a local 
enhancement in density near the Earth, compared to the halo WIMP density:

\begin{equation}
\delta_E \equiv \frac{n (a_1)}{n_X}  \
= \frac{0.21}{(v_o / 220 \, 
{\rm kms}^{-1})} \, g_{\rm tot}^{(-10)} \, , 
\end{equation}
where $g_{\rm tot}^{(-10)} \equiv 10^{10} \, g_{\rm tot} ({\rm GeV})^3$ 
(i.e.,for masses given in GeV and cross sections in units of 
$10^{-10} \, {\rm GeV}^{-2}$, with $\hbar = c =1$).

The a priori reasonableness of the final
density enhancement (5) can be seen as follows.
From Eqs. (1) and (2)  the total rate (for
$J_{\min} = 0$) of WIMPs scattered within 1 AU
is of order $v_E^2 d\dot N/d \alpha \sim
n_X m_S g_{\rm tot} v_E^2/v_o$. Integrating over
the lifetime $t_S$  of the solar system
and dividing by the number of galactic WIMPs
present within 1AU gives a maximum possible density enhancement
$\delta_{\rm max} \sim (3/4\pi)(m_S/a_E^3)(v_E^2/v_o) t_S g_{\rm tot} 
\sim 180 g_{\rm tot}^{(-10)}$.
Out of this, detailed calculations of planetary perturbations of
the WIMP's captured in the Sun's outer regions show \cite{damkrauss}
that a fraction {\it parametrically} of order 
$(\epsilon_m)^{0.4} (a_1/R_S)^{0.3} ({\sum_{p}} \mu_p (a_1/a_p)^3)^{0.6}
\sim 10^{-3}$ is extracted. (Here $\epsilon_m$=0.022 accounts for the
density decrease of the Sun's outer regions of interest here.)

Perhaps a more relevant quantity, from the point of view of experiment, is
the differential rate, $dR / dQ$, per keV per kg per day, of scattering events
in a laboratory sample made of element $A$.  Comparing this differential rate
due to our new WIMP population with that due to the parent galactic
halo population we find a ratio which reaches the flat maximum,

\begin{equation}
\rho (Q) \equiv \frac{(dR / dQ)^{\rm new}}{(dR / dQ)^{\rm standard}}
\  = 1.1 \, g_{\rm tot}^{(-10)} \, ,
\end{equation}
for energy transfers $Q$ smaller than
$Q_E = 2 (m_X/(m_X + m_A))^2 m_A v_E^2$. For a target of Germanium
(the present material of choice in several ongoing cryogenic
WIMP detection experiments), $Q_E$ is in the keV range.
If $g_{\rm tot}^{(-10)} \sim 1$  this yields a 100~\% increase of 
the differential event rate below the value $Q = Q_E \sim {\rm keV}$ which is
typical for the energy deposit due to the new WIMP population, whose
characteristic velocity $ v_E$ is $\sim 7$ times smaller than that
of galactic halo WIMPs.

\begin{figure}[tbp]
\epsfxsize = \hsize \epsfbox{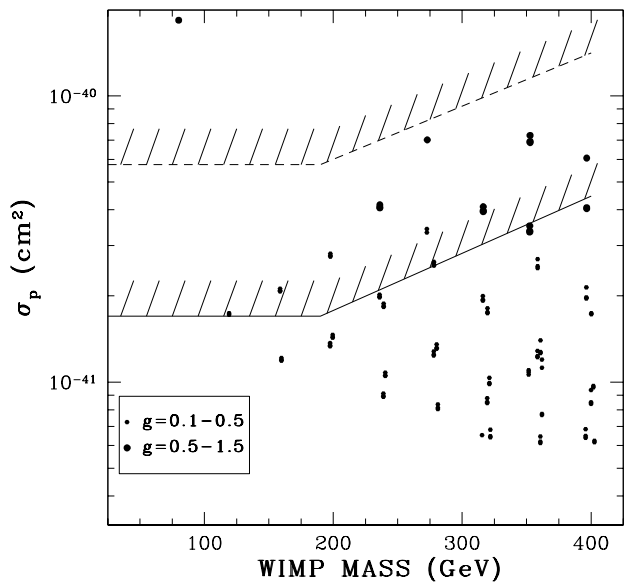}
\caption{Values of $g_{\rm tot}^{(-10)}$ as a function of WIMP nucleon 
cross section and WIMP mass for a sample range of allowed SUSY
models ($\mu$ parameter $ >0$) . Also shown are approximate
experimental upper limits on WIMP nucleon cross section from direct
detection experiments [13], assuming two different values for the local
galactic halo dark matter density.($\rho = 0.3 {\rm GeV cm^{-3}}$ (lower)
and $\rho = 0.1 {\rm GeV cm^{-3}}$ (upper) } 
\end{figure}

This is our central result. Its relevance
depends completely on the actual value of $g_{\rm 
tot}^{(-10)}$.  As an example of the possibilities, 
in Figure 1 we display the
range of
$g_{\rm tot}$ derived by sampling over the allowed Minimal Supersymmetric
Standard Model phase space (values for SUSY
parameter $\mu > 0$ are chosen here), and for WIMP density in the range 
$.05 h^{-2}< \rho_X/\rho_{\rm closure} <1$ 
(where $h$ is the Hubble parameter in units of
 $100 {\rm km s^{-1}Mpc^{-1}} $). 
 Values of $g_{\rm tot}$ in excess of 1 are thus possible.

 Existing detectors tend to
lose sensitivity in the range of a few keV.  However, our result provides
motivation to consider pushing hard in this direction. The
new population will have a strongly anisotropic velocity distribution.  Not
only might this greatly help distinguish it from backgrounds, but a
comparison of the angular anisotropy and annual modulation of 
this distribution with the corresponding features of the higher energy signal 
from a halo WIMP distribution would be striking.  
We emphasize once again that if neutralinos exist in the range detectable at
the next generation of detectors, this new WIMP population should exist at
detectable levels as well.  Finally, we note that the indirect neutrino 
signature
of such WIMPs that might subsequently be captured by the Earth, and annihilate,
could be dramatic.  The new population has a characteristic velocity that more
closely matches the escape velocity from the Earth than does the background halo
population.  As a result, resonant capture off elements such as iron in the
Earth could be greatly enhanced.

This work was supported in part by the U.S. DOE.  LMK acknowledges
the  hospitality of the IHES, where much of this work was carried out, and
CERN, where this work was initiated. We thank D. Devaty and P. Kernan for
assistance with various numerical routines.

\end{document}